\documentclass[11pt,a4paper]{scrartcl}
\usepackage[utf8]{inputenc}
\usepackage[T1]{fontenc}
\usepackage{lmodern}

\usepackage{graphicx}
\usepackage{epsfig}
\usepackage{float}
\usepackage{cite}
\usepackage{amssymb}
\usepackage{amsmath}
\usepackage{url}
\usepackage[dvipsnames]{xcolor}
\usepackage{cancel}
\usepackage{nicefrac}
\usepackage[normalem]{ulem}
\usepackage[toc,page]{appendix}
\usepackage[onehalfspacing]{setspace}
\usepackage{booktabs}

\usepackage[format=plain,labelfont={bf},font={small}]{caption}

\usepackage{hyperref}

\usepackage{comment}

\setkomafont{disposition}{\normalcolor\normalfont\bfseries}

\newcommand*{\alphas}{\alpha_{\text{s}}}

\newcommand*{\mr}{\mathrm}
\newcommand*{\mc}{\mathcal}

\newcommand*{\beq}{\begin{equation}}
\newcommand*{\eeq}{\end{equation}}
\newcommand*{\bea}{\begin{eqnarray}}
\newcommand*{\eea}{\end{eqnarray}}

\newcommand*{\mll}{M_{\ell^+\ell^-}} 
\newcommand*{\mee}{M_{e^+e^-}} 
 
\newcommand*{\muf}{\mu_\mr{F}} 
\newcommand*{\mur}{\mu_\mr{R}} 
\newcommand*{\xif}{\xi_\mr{F}} 
\newcommand*{\xir}{\xi_\mr{R}} 
\newcommand*{\ee}{e^+e^-} 
\newcommand*{\mumu}{\mu^+\mu^-}

\newcommand*{\tth}{$t\bar t H$}

\newcommand*{\ttz}{$t\bar t Z$} 
\newcommand*{\ttw}{$t\bar t W$} 
\newcommand*{\ttwpm}{$t\bar t W^\pm$}

\newcommand*{\ttll}{$t\bar t \ell^+\ell^-$} 
\newcommand*{\ttee}{$t\bar t e^+e^-$}

\newcommand*{\ttzdec}{(t\to\mu^+\nu_\mu b) (\bar t\to\mu^-\bar\nu_{\mu} \bar b) \, \ee}

\newcommand*{\ttzdecz}{(t\to\mu^+\nu_\mu b) (\bar t\to\mu^-\bar\nu_{\mu} \bar b) \, (Z\to e^+ e^-)}

\newcommand*{\ttepem}{t\bar t e^+e^-} 
\newcommand*{\ppttz}{pp\to t\bar t Z}
\newcommand*{\ppttw}{pp\to t\bar t W^\pm} 
 
\newcommand*{\ppttll}{pp\to t\bar t \ell^+\ell^-} 
\newcommand*{\ppttee}{pp\to t\bar t e^+e^-} 

\newcommand*{\dytt}{\Delta y_{t\bar t}} 
\newcommand*{\mtt}{M_{t\bar t}} 
\newcommand*{\ttbar}{t\bar t}

\newcommand*{\POWHEGBOX}{\texttt{POWHEG BOX}} 
\newcommand*{\POWHEGBOXV}{\texttt{POWHEG BOX-V2}} 

\newcommand*{\POWHEG}{POWHEG} 
\newcommand*{\MCNLO}{MC@NLO}
\newcommand*{\HERWIG}{\texttt{HERWIG}} 
\newcommand*{\PYTHIA}{\texttt{PYTHIA}}

\newcommand*{\Madgraph}{\texttt{MadGraph}} 
\newcommand*{\NLOX}{\texttt{NLOX}} 

\usepackage{comment}

\begin{document}

\renewcommand*{\thefootnote}{\fnsymbol{footnote}}

\begin{center}
	{
\Large \textbf{Hadronic production of top-quark pairs in association
  with a pair of leptons in the POWHEG BOX framework}}\\
	\vspace{.7cm}
	Margherita Ghezzi\footnote{\texttt{margherita.ghezzi@itp.uni-tuebingen.de}},
	Barbara J\"ager\footnote{\texttt{jaeger@itp.uni-tuebingen.de}}, 
	Santiago Lopez Portillo Chavez\footnote{\texttt{santiago.lopez-portillo-chavez@uni-tuebingen.de}}, 
	\\
%	\vspace{.3cm}
	\textit{
		Institute for Theoretical Physics, University of T\"ubingen, Auf der Morgenstelle 14, 72076 T\"ubingen, Germany
	}
	\\[.3cm]		 
	Laura Reina\footnote{\texttt{reina@hep.fsu.edu}}, 
	\\
%	\vspace{.3cm}
	\textit{
		Physics Department, Florida State University, Tallahassee, FL 32306-4350, USA
	}
	\\[.3cm]
	Doreen Wackeroth\footnote{\texttt{dw24@buffalo.edu}}, 
	\\
	\textit{
		Department of Physics, University at Buffalo, The
                State University of New York, \\239 Fronczak Hall, Buffalo, NY 14221, USA
	}
\end{center}   

\renewcommand*{\thefootnote}{\arabic{footnote}}
\setcounter{footnote}{0}

\vspace*{0.1cm}
\begin{abstract}
  We present an implementation of \ttll{} ($\ell=e,\mu$) hadronic production
  at next-to-leading order in QCD matched to parton-shower event generators
  in the \POWHEGBOX{} framework. The program we developed includes all
  leading-order contributions of order $\alpha_s^2\alpha^2$ for the
  specified final state, as well as the corresponding first-order QCD
  corrections. Decays of the top quarks have been simulated retaining
  spin-correlations in all tree-level matrix elements. We consider the
  case of the Large Hadron Collider at $\sqrt{s}=13$~TeV and compare
  results for \ttll{} production in the fiducial volume where the
  invariant mass of the lepton pairs is centered around the $Z$-boson
  mass to corresponding predictions for $t\bar{t}Z$ on-shell
  production with $Z\rightarrow \ell^+\ell^-$. We find that off-shell
  effects in \ttll{} are in general small at the level of the total
  cross section, but can decrease the tail of the leptons' transverse
  momentum distributions by 10-20\% and, in these regions, they are
  visible beyond the scale uncertainty due to renormalization and
  factorization scale variation. Moreover, we find that accounting for
  top-quark decays in the narrow-width approximation with
  tree-level spin correlations also gives origin to 10-20\% effects in
  specific regions of the kinematic distributions of the \ttll{} decayed final state.
\end{abstract}
\newpage

%\tableofcontents

%%%%%%%%%%%%%%%%%%%%%%%%%%%%%%%%%%%%%%%%%%%
\section{Introduction}\label{sec:intro}
%%%%%%%%%%%%%%%%%%%%%%%%%%%%%%%%%%%%%%%%%%%
%
The CERN Large Hadron Collider (LHC) provides energies and
luminosities high enough to produce an unprecedented number of top
quarks.  As a top-quark factory, it constitutes an ideal environment
for exploring the properties of the heaviest known elementary fermion
to date. While several intrinsic properties of the top quark are
accessible in top-quark pair production processes ($pp\rightarrow
t\bar{t}$), its couplings to the electroweak $W$ and $Z$ bosons can
best be probed in associated production processes, such as $\ppttz$
and $\ppttw$.  Deviations of measured production rates and kinematic
distributions from Standard Model (SM) predictions can provide hints
of new physics effects that can then be interpreted in terms of
specific models or more general effective interactions.  At the same
time, the production of \ttwpm{} and \ttz{} are important backgrounds
in the measurement of the associated production of the Higgs boson
with a top quark pair ($pp\rightarrow t\bar{t}H$) as well as more
generally in searches of new physics which involve final states with
multiple leptons and $b$ jets.

Cross sections for the \ttwpm{} and \ttz{} production processes
measured at the LHC with center-of-mass energy $\sqrt{s} = 13$~GeV by
the ATLAS~\cite{ATLAS:2016wgc,ATLAS:2019fwo,ATLAS:2021fzm} and CMS
experiments~\cite{CMS:2017ugv,CMS-PAS-TOP-18-009} tend to be in
agreement with SM expectations, but are plagued by relatively large
statistical and systematic uncertainties. A crucially limiting factor
in the analyses is due to theoretical uncertainties. The expected
improvement of experimental accuracy in the upcoming LHC Run~3 and the
future LHC high-luminosity run (HL-LHC) will unfold its full potential
only if accompanied by a corresponding improvement in the precision of
theoretical predictions and simulation tools.

Providing the means to identify effects of new physics in deviations
of measurements from SM expectations not only requires precise
theoretical predictions for benchmark observables in the context of
the SM and representative scenarios for its extensions, but also a
realistic assessment of systematic uncertainties in experimental
analyses and theoretical calculations. Such uncertainties include
unknown perturbative corrections (beyond the orders in the strong and
electroweak couplings considered in a calculation), non-perturbative
effects (which are typically estimated using some generic models), and
intrinsic uncertainties of the simulation programs used. In particular
the latter can only be quantified by a comparison of conceptually
different implementations.

Several precise predictions and simulation tools for the hadronic
production of top-quark pairs in association with weak gauge bosons
are available.  Results for the corresponding total and differential
cross sections including next-to-leading order (NLO) QCD and
electroweak (EW) corrections have been presented
in~\cite{Campbell:2012dh,Garzelli:2012bn,Maltoni:2015ena,Frixione:2015zaa,Frederix:2017wme,Frederix:2020jzp}
for \ttwpm{} and
in~\cite{Lazopoulos:2008de,Kardos:2011na,Maltoni:2015ena,Frixione:2015zaa}
for \ttz, respectively.\footnote{A review and comparison of
  fixed-order results has also been presented in the 4$^{\mathrm{th}}$
  report of the LHC Higgs Cross Section Working
  Group~\cite{LHCHiggsCrossSectionWorkingGroup:2016ypw}.} Soft-gluon
resummation effects up to next-to-next-to-leading logarithms (NNLL)
have been studied
in~\cite{Li:2014ula,Broggio:2016zgg,Kulesza:2018tqz,Broggio:2019ewu,Kulesza:2020nfh,vonBuddenbrock:2020ter}.
Furthermore, the modeling of \ttwpm{} and \ttz{} events have been
improved by interfacing fixed-order NLO QCD calculations with parton
showers (PS) in several studies
~\cite{Garzelli:2012bn,Garzelli:2011is,Maltoni:2014zpa,Frederix:2020jzp,Frederix:2021agh,Cordero:2021iau}
based either on the POWHEG~\cite{Nason:2004rx,Frixione:2007vw} or
\MCNLO~\cite{Frixione:2002ik,Frixione:2003ei} methods.  Some of these
frameworks also provide extra functionalities such as the simulation
of top-quark decays including spin correlations at leading order (LO)
accuracy~\cite{Maltoni:2014zpa,Frederix:2020jzp,Frederix:2021agh,Cordero:2021iau}
and a systematic treatment of higher jet multiplicities via multi-jet
merging techniques~\cite{Frederix:2020jzp,Frederix:2021agh}.

Calculations also exist which take NLO QCD effects and off-shell
effects in the decays of the top quarks and/or $W^\pm$ bosons into
account~\cite{Campbell:2012dh,Bevilacqua:2020pzy,Bevilacqua:2020srb,Denner:2020hgg,Denner:2021hqi}.
While fixed-order NLO QCD corrections in both production and decay of
the top quarks and $Z$~boson have been taken into account for instance
in \cite{Rontsch:2014cca,Bevilacqua:2019cvp,Hermann:2021xvs}, a full
study of off-shell effects in the $Z$ boson's decay modes to charged
fermions is still missing.  It is our goal in this paper to study the
$\ppttll$ process (with $\ell^\pm$ denoting either $e^\pm$ or
$\mu^\pm$) including both resonant and non-resonant $Z$- and
photon-induced contributions, and explore their impact on
experimentally accessible quantities such as angular correlations of
the decay leptons $\ell^+$ and $\ell^-$. Our study includes all
LO contributions of order $\alpha_s^2\alpha^2$ for the
specified final state, as well as the corresponding first-order QCD
corrections. Moreover, we aim at presenting a dedicated Monte Carlo
program allowing not only the calculation of NLO QCD corrections to
the $\ppttll$ production process, but also providing the option of
taking into account decays of the top quarks with full tree-level spin
correlations in the narrow-width approximation, using the method of
Ref.~\cite{Frixione:2007zp}.

Furthermore we have implemented the NLO QCD calculation of this
process in \POWHEGBOX{}~\cite{Alioli:2010xd}, which provides an
interface to general-purpose Monte Carlo PS event generators such as
\PYTHIA{} and \HERWIG{} using the \POWHEG{}
method~\cite{Nason:2004rx,Frixione:2007vw}.  For completeness, we have
also developed an independent implementation of $\ppttz${} at NLO QCD
matched to PS using the \POWHEG{} method in \POWHEGBOX, and will
describe the implementation of both processes in the course of this
paper. We notice that for the first time the implementation of both
processes will be available as part of the \POWHEGBOXV{} repository,
which also includes processes such as $pp\rightarrow
t\bar{t}H$~\cite{Hartanto:2015uka,Jager:2015hka}. This will allow to
generate both signal and background events in a consistent framework
for studies of Higgs-boson associated production with top-quark
pairs. We hope that the tools described in this paper will help in the
analysis and interpretation of data collected by the ATLAS and CMS
experiments in upcoming LHC runs and will facilitate the study of
residual modeling
uncertainties~\cite{ATL-PHYS-PUB-2020-024,Bevilacqua:2021tzp}.

This article is structured as follows. In the next section we explain
the implementation of the \ttll production process in the \POWHEGBOX{}
framework, while detailed numerical studies are presented in
Sec.~\ref{sec:results}. We summarize our findings in
Sec.~\ref{sec:conclusion}.

%%%%%%%%%%%%%%%%%%%%%%%%%%%%%%%%%%%%%%%%%%%
\section{Implementation}\label{sec:implementation}
\begin{figure}\centering
\scalebox{0.70}{
\begin{minipage}{0.55\textwidth}
		\centering
		\includegraphics[width=0.9\textwidth]{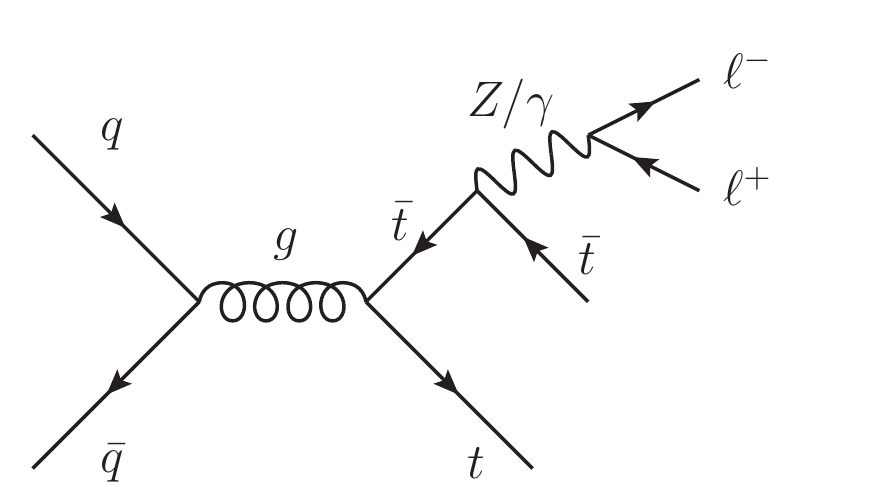}
\end{minipage}
}
\scalebox{0.70}{
\begin{minipage}{0.55\textwidth}
	\centering
	\includegraphics[width=0.8\textwidth]{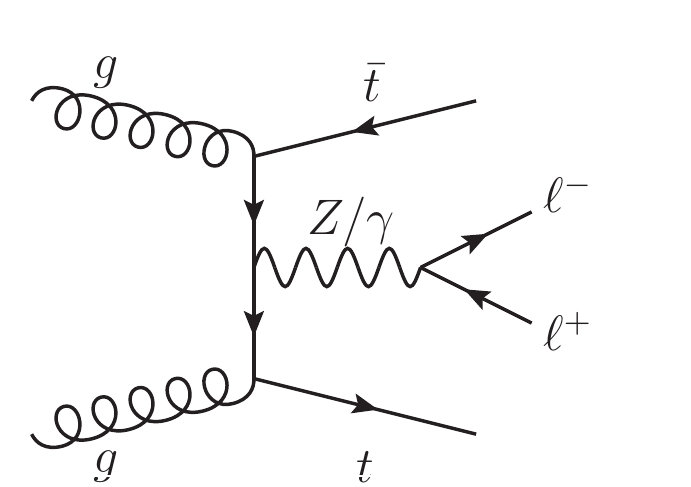}
\end{minipage}
}
\scalebox{0.70}{
\begin{minipage}{0.55\textwidth}
	\centering
	\includegraphics[width=0.9\textwidth]{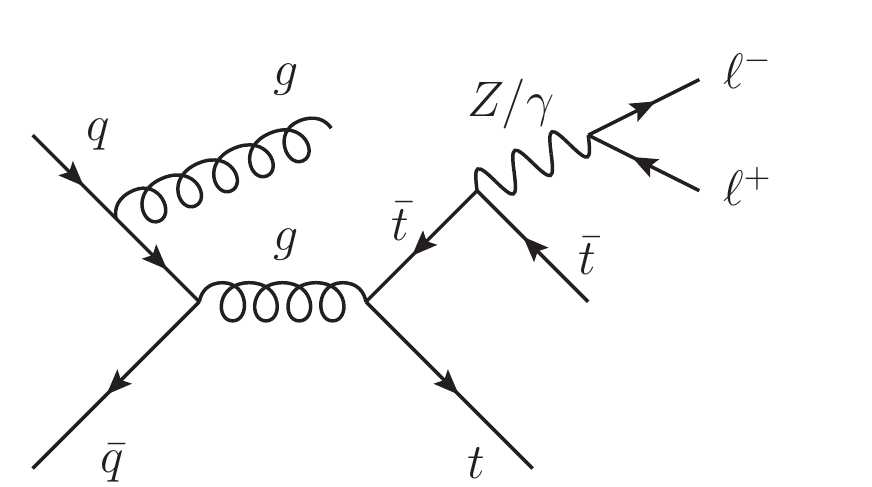}
\end{minipage}
}
\scalebox{0.70}{
\begin{minipage}{0.55\textwidth}
	\centering
	\includegraphics[width=0.8\textwidth]{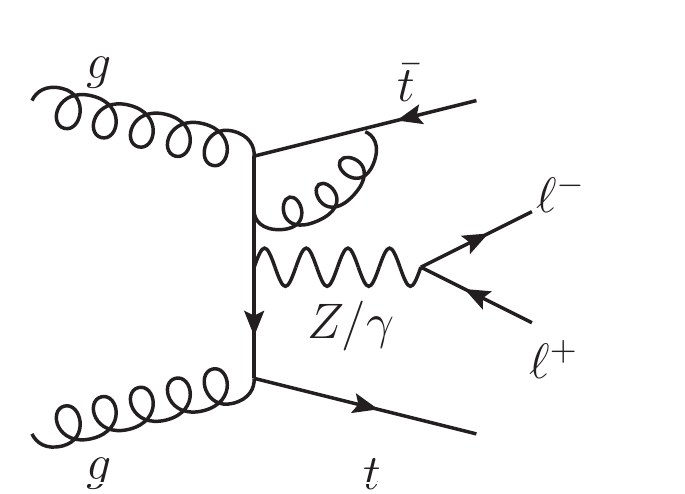}
\end{minipage}
}
\caption{Representative Feynman diagrams for quark- and gluon-induced
  contributions to the $pp \to t\bar{t} \ell^+\ell^-$ process. The top
  row illustrates examples of processes that contribute to the LO QCD
  cross section at order $\alphas^2\alpha^2$, while the bottom row
  provides examples of real-emission (bottom left) and virtual (bottom
  right) NLO QCD corrections.  }
\label{fig:rep_FD}
\end{figure}

%%%%%%%%%%%%%%%%%%%%%%%%%%%%%%%%%%%%%%%%%%%
%
We consider the production of the \ttll{} final state in proton-proton collisions ($\ppttll$, with $\ell=e,\mu$) and calculate the corresponding cross section up to NLO in the strong coupling $\alphas$ and at LO in the EW coupling $\alpha$, i.e. we consider LO contributions of order $\alpha_s^2\alpha^2$ and NLO QCD contributions of order $\alpha_s^3\alpha^2$. 
%all terms of the perturbative expansion up to $\mc{O}(\alphas^3\alpha^2)$ are retained. 
Since we assume both electrons and muons to be massless, the calculation for both cases is the same. Representative diagrams are shown in Fig.~\ref{fig:rep_FD} for both LO contributions (top row) and real and virtual NLO QCD corrections (bottom row).
%
%\begin{figure}\centering
%\scalebox{0.70}{
%\begin{minipage}{0.55\textwidth}
%\centering
%\input{figures/FD/qq1}
%\end{minipage}
%}
%\scalebox{0.70}{
%\begin{minipage}{0.55\textwidth}
%\centering
%\input{figures/FD/gg2}
%\end{minipage}
%}
%\scalebox{0.70}{
%\begin{minipage}{0.55\textwidth}
%\centering
%\input{figures/FD/qq1_real}
%\end{minipage}
%}
%\scalebox{0.70}{
%\begin{minipage}{0.55\textwidth}
%\centering
%\input{figures/FD/gg2_virt}
%\end{minipage}
%}
As part of our study we also consider the corresponding production of
an on-shell \ttz{} final state ($\ppttz$) followed by the $Z$-boson
decaying into a pair of charged leptons ($Z\rightarrow \ell^+\ell^-$)
and calculate it at the same perturbative order
(i.e. $\mc{O}(\alphas^3\alpha)$ for \ttz{} production and
$\mc{O}(\alphas^3\alpha^2)$ when including the decay of the
$Z$-boson).\footnote{In order to test our $\ppttz${} implementation,
  we compared results obtained with our code for \ttz{} production to
  results presented in
  Ref.~\cite{LHCHiggsCrossSectionWorkingGroup:2016ypw}. Using the
  setup of Ref.~\cite{LHCHiggsCrossSectionWorkingGroup:2016ypw}, we
  find excellent agreement at fixed-order NLO QCD for both total and
  differential rates.}

For the implementation of the $\ppttll$ and $\ppttz${} processes in
the framework of the {\texttt V2} version of \POWHEGBOX{}, we resort
to a combination of existing tools and tailored building blocks. While
\POWHEGBOX{} provides all process-independent elements of the matching
of an NLO QCD calculation to parton-shower event generators, it
requires process-specific input from the developer such as the flavor
structure of the given process, a parameterization of its phase space,
the hard-scattering matrix elements squared at LO, the NLO virtual
corrections and real-emission amplitudes squared, as well as spin- and
color-correlated amplitudes for the construction of infrared
subtraction terms according to the FKS scheme~\cite{Frixione:1995ms}.

In order to build the required tree-level amplitudes, subtraction
terms, and flavor structure for the $\ppttll${} and $\ppttz${}
processes, we rely on standard features of \POWHEGBOX{} based on
\Madgraph~4~\cite{Alwall:2007st}.  We provide the required \ttll{}
phase-space parametrization with the option of mapping the $Z$-boson
resonance if the selected fiducial volume of a given study is limited
to events where the invariant mass of a pair of opposite-sign
same-flavor leptons is restricted to a narrow region around the
$Z$-boson mass.  Virtual corrections are computed with the help of the
one-loop provider
\NLOX{}~\cite{Honeywell:2018fcl,Figueroa:2021txg}. The interface that
ensures the correct calls of the \NLOX{} amplitudes within
\POWHEGBOX{}, as well as the transfer of the necessary input
parameters, has also been developed.

While the cross section for the \ttz{} production process is finite at
Born level without any phase-space restrictions, the cross section for
the \ttll{} final state exhibits a collinear singularity already at
tree level because of a photon of vanishing virtuality splitting into
an $\ell^+\ell^-$ pair of massless leptons, c.f.\ the photon-exchange
diagrams in Fig.~\ref{fig:rep_FD}.  Experimentally, such
configurations are of little interest since measurements are typically
performed by imposing a cut on the invariant mass of the lepton pair
($\mll$) that favors configurations where the lepton pair originates
from a decaying $Z$ boson (rather than a photon of low virtuality). To
exclude this singular phase-space region, we implemented a generation
cut on the invariant mass of the $\ell^+ \ell^-$ pair,
$\mll^\mr{min}$, which is imposed at the level of the Born phase-space
generation. Consequently, any analysis cut on the invariant mass of
the $\ell^+ \ell^-$ system
%imposed at a later stage of the event generation
has to be chosen larger than $\mll^\mr{min}$.  The results
presented in Sec.~\ref{sec:results} have been obtained with
$\mll^\mr{min}=10$~GeV and an analysis cut of
$m_Z-10~\mr{GeV}\leq\mll\leq m_Z+10$~GeV.

The most basic version of our implementation treats the top quarks as
stable particles. Decays of the top quarks, for instance via the chain
$t\to W^+b\to \ell'^+\nu_{\ell'} b$, can in principle be simulated in
the narrow-width approximation after the event-generation stage by
using the decay feature of a multi-purpose Monte Carlo event generator
like \PYTHIA.  However, such a simulation of the top-quark decays is
limited to the spin-averaged case and cannot provide information on
correlations between the production and decay part of the full $pp\to
( \ell'^+\nu_{\ell'} b)( \ell'^-\bar\nu_{\ell'} b)\ell^+\ell^-$
process.

In order to partially overcome these limitations, we also introduce
the possibility of retaining the correlations between production and
decays of the top quarks, in the narrow-width approximation, applying
the method of Ref.~\cite{Frixione:2007zp}. Implementations of this
method in the context of \POWHEGBOX{} can be found in related
processes such as $t\bar t +$jet and \tth{} production
\cite{Alioli:2011as,Hartanto:2015uka}. Inspired by these examples, we
proceed as follows. In a first step, the program internally generates
events at NLO QCD accuracy for \ttll{} final states with stable top
quarks. Subsequently, decays of the top quarks into bottom quarks,
leptons, and neutrinos are simulated according to a probability
determined by tree-level matrix elements for $(t\to \ell^{\prime
  +}\nu_{\ell^\prime} b)(\bar t\to \ell^{\prime
  -}\bar\nu_{\ell^\prime} \bar b)\, \ell^+\ell^- (+$jet)
production. The corresponding matrix elements have been obtained with
\Madgraph~5~\cite{Alwall:2014hca}. In this way, the full tree-level
spin correlations between production and decay are retained in the
soft and collinear regions as well as in the real-emission matrix
elements. For more details on the method and its implementation in
\POWHEGBOX{}, we refer the interested reader to the original
publications of Refs.~\cite{Frixione:2007zp,Alioli:2011as}.

%%%%%%%%%%%%%%%%%%%%%%%%%%%%%%%%%%%%%%%%%%%
\section{Numerical analysis and results}\label{sec:results}
%%%%%%%%%%%%%%%%%%%%%%%%%%%%%%%%%%%%%%%%%%%
%
In the following, we explore the capabilities of the program we
developed and present some representative phenomenological results for
the LHC running at a center-of-mass energy of 13~TeV.  For
illustration purposes, we specialize our discussion to the case of
$pp\to t\bar t e^+e^-$ and consider the top and anti-top decays
$t\rightarrow \mu^+\nu_\mu b$ and
$\bar{t}\rightarrow \mu^-\bar{\nu}_\mu \bar{b}$. As explained in
Sec.~\ref{sec:implementation}, our \POWHEGBOX{} implementation also
allows to consider other choices for the leptons coming from \ttll{}
production and top/antitop decays, with the obvious caveat that the
corresponding selection of events (i.e. the selection of the
final-state fiducial volume via a set of cuts on final-state
particles) needs to be modified accordingly.

The effect of NLO QCD corrections on \ttee{} production will be
presented in Sec.~\ref{sec:ttee-nlo-qcd} for both total and
differential distributions, considering a variety of kinematic
observables. In particular, we will discuss the charge asymmetry of
the $t\bar{t}$ system. In order to identify the observables that are
most affected by off-shell effects and the interplay between $Z$ and
photon resonant channels in \ttee{} production, in
Sec.~\ref{sec:results-off-shell} we will compare our findings to the
corresponding results obtained from \ttz{} on-shell production
matched, in the narrow-width approximation, to the decay of the $Z$
boson into an $\ee$ pair as simulated by the decay feature of
\PYTHIA{}, the PS Monte Carlo event generator adopted in our study,
where spin-correlation effects are not taken into account. % In this
%comparison the decays of the on-shell top and antitop quark will be
%implemented using a narrow-width approximation, and with the option of
%including or not including spin-correlation effects for all tree-level
%matrix elements, as discussed in Sec.~\ref{sec:implementation}. A
%comparison between the two treatments will be presented in
%Sec.~\ref{sec:results-top-decays}.
In Sec.~\ref{sec:results-top-decays} we present an assessment of the 
effect of spin correlations in the $\ppttee{}$ process, where we
compare the two options of top-decay modeling available in our implementation. 
We end the section by comparing our most complete
modeling of the process, $\ppttee{}$ with spin-correlated top decays,
against the simplest one, $pp\to t\bar{t}(Z\to e^+ e^-)$ without spin-correlated top decays.

The results presented in this section have been obtained using the
CT18NLO set~\cite{Hou:2019efy} of parton distribution functions (PDF)
as implemented in the LHAPDF6 library~\cite{Buckley:2014ana},
corresponding to $\alpha_s(m_Z)=0.118$ for five massless quarks.  For
the EW input parameters we use the $G_\mu$ scheme where, besides the
Fermi constant $G_\mu$, the masses of the $Z$ and $W$ bosons are
fixed. For our study we choose the following input
values~\cite{ParticleDataGroup:2020ssz}:
\begin{equation}
m_Z = 91.1876~\mr{GeV}\,,\quad m_W =
80.379~\mr{GeV}\,,\quad G_\mu = 1.166378\times
10^{-5}~\mr{GeV}^{-2}\,.  
\end{equation}
Other EW parameters like the EW
coupling $\alpha$ and the weak mixing angle $\theta_W$ are computed
thereof via tree-level relations. The widths of the $Z$ and $W$ bosons
are set to~\cite{ParticleDataGroup:2020ssz}: 
\begin{equation}
\Gamma_Z= 2.4952~\mr{GeV}\,,\quad \Gamma_W=
2.085~\mr{GeV}\,,
\end{equation}
while for the top-quark mass and width we use~\cite{ParticleDataGroup:2020ssz}: 
\begin{equation}
m_t= 172.76~\mr{GeV}\,, \quad
\Gamma_t= 1.42~\mr{GeV}\,.
\end{equation}
We define the renormalization and factorization scales as multiples of
a central scale $\mu_0$ according to $\mur=\xir \mu_0$ and
$\muf=\xif\mu_0$ with variation factors $\xi_R$ and $\xi_F$, and
consider both the case of a fixed central scale: \beq
\label{eq:fixed-scale}
\mu_0 = \frac{2m_t+m_Z}{2}\,,
\eeq
and the case of a dynamical central scale:
\begin{equation}
\label{eq:dyn-scale}
 \mu_0 = \frac{M_T(e^+e^-) + M_T(t) + M_T(\bar{t})}{3}\,, 
\end{equation}
based on the transverse masses of the top quark and anti-quark, and of
the $\ee$ system, where the transverse mass $M_T(i)$ of a particle or
system of particles $i$ is obtained from its (invariant) mass $m_i$
and transverse momentum, $p_{T,i}$, via
\begin{equation}
M_T(i)=\sqrt{m_i^2+p_{T,i}^2}\,.
\end{equation}
In our \ttee implementation, $M_T(e^+e^-)$ is computed from the
momenta of the $\ee$~system, while in the \ttz{} code it is determined
by the momentum and mass of the $Z$~boson. All results in this paper
include a renormalization and factorization scale uncertainty obtained
by a seven-point scale variation, i.e.~by independently setting the
scale factors $\xif$ and $\xir$ to the values $\frac{1}{2}$, $1$ and
$2$ while excluding the combinations $(\xif,\xir) = (\frac{1}{2},2)$
and $(\xif,\xir) = (2,\frac{1}{2})$. As we will see in
Sec.~\ref{sec:ttee-nlo-qcd}, the two scale choices give very
similar results and in presenting the results of
Secs.~\ref{sec:results-off-shell} and \ref{sec:results-top-decays}
we will consider only the fixed scale choice of
Eq.~(\ref{eq:fixed-scale}), unless otherwise specified.  In estimating the theoretical uncertainty
of the results presented in this paper we do not include PDF
uncertainties since, based on studies of \ttz{} production (see
e.g. ~\cite{LHCHiggsCrossSectionWorkingGroup:2016ypw}), they are at
the moment subleading compared to residual uncertainties from scale
variation.

Our \POWHEGBOX{} implementations provide event files in the LHA
format~\cite{Alwall:2006yp} that we subsequently process with
\PYTHIA~\texttt{8.240}~\cite{Sjostrand:2014zea} using the
\texttt{Monash 2013} tune~\cite{Skands:2014pea}. For our simulations
we deactivate multi-parton interactions (MPI) and hadronization effects and switch off QED
showering.

If not specified otherwise, in our phenomenological analyses cuts are
imposed on the transverse momentum and pseudorapidity of the electron
and positron of the \ttee final state, 
\begin{equation}
\label{eq:ptl-cut}
p_T^{e} > 10~\mr{GeV}\,,\quad
|\eta^e| < 2.5\,,\quad
\end{equation}
and on the invariant mass of the $\ee$~system, restricting it to a
window around the $Z$-boson mass, i.e.:
\begin{equation}
\label{eq:mee-cut}
m_Z-10~\mr{GeV}\leq\mee\leq m_Z+10~\mr{GeV}\,.
\end{equation}
When decays of the top and anti-top quarks are taken into account,
namely $t\to \mu^+\nu_\mu b$ and $\bar t\to \mu^-\bar\nu_\mu \bar b$,
the following additional cuts on the muons' transverse momenta and
pseudorapidity are also imposed, 
\begin{equation}
\label{eq:muon-cuts}
p_T^{\mu} > 10~\mr{GeV}\,,\quad
|\eta^\mu| < 2.5\,.\quad
\end{equation}

%%%%%%%%%%%%%%%%
\subsection{Features of the NLO QCD corrections to \ttll{} production }
\label{sec:ttee-nlo-qcd}
In order to assess the theoretical uncertainty of our fixed-order
predictions, we consider total and differential cross sections for
$\ppttee$ after the lepton cuts of Eqs.~(\ref{eq:ptl-cut}) and
(\ref{eq:mee-cut}) have been applied and investigate the impact of NLO
QCD corrections and their dependence on the renormalization and
factorization scales.  First of all we notice that total cross
sections show a very mild dependence on the nature of the
renormalization and factorization scales. Choosing the fixed scale of
Eq.~(\ref{eq:fixed-scale}) we obtain the following LO and NLO QCD
cross sections:
\begin{eqnarray}
\label{eq:lo-nlo-fixed-scale}
 \sigma_\mr{\ttepem}^\mr{LO} &=& 15.9^{+5.1}_{-3.6}  \, \mr{fb} \; ,\\ \nonumber
\sigma_\mr{\ttepem}^\mr{NLO} &=& 21.9^{+2.0}_{-2.4}  \, \mr{fb} \; ,
\end{eqnarray}
while the corresponding results when using the dynamical scale of Eq.~(\ref{eq:dyn-scale}) are
\begin{eqnarray}
\label{eq:lo-nlo-dynamical-scale}
\sigma_\mr{\ttepem}^\mr{LO} &=& 15.8^{+5.0}_{-3.5}  \, \mr{fb} \; , \\ \nonumber
\sigma_\mr{\ttepem}^\mr{NLO} &=& 22.1^{+2.2}_{-2.5}  \, \mr{fb} \; ,
\end{eqnarray}
where LO and NLO QCD results have been calculated for the same choice
of NLO PDF and $\alpha_s$ while the quoted uncertainties are from
renormalization and factorization scale variation, as discussed in the beginning of Sec.~\ref{sec:results}.
%%%%%%%%%
%
%\begin{figure}[htb]
\begin{figure}[t]
	\centering
	\includegraphics[width=0.48\textwidth]{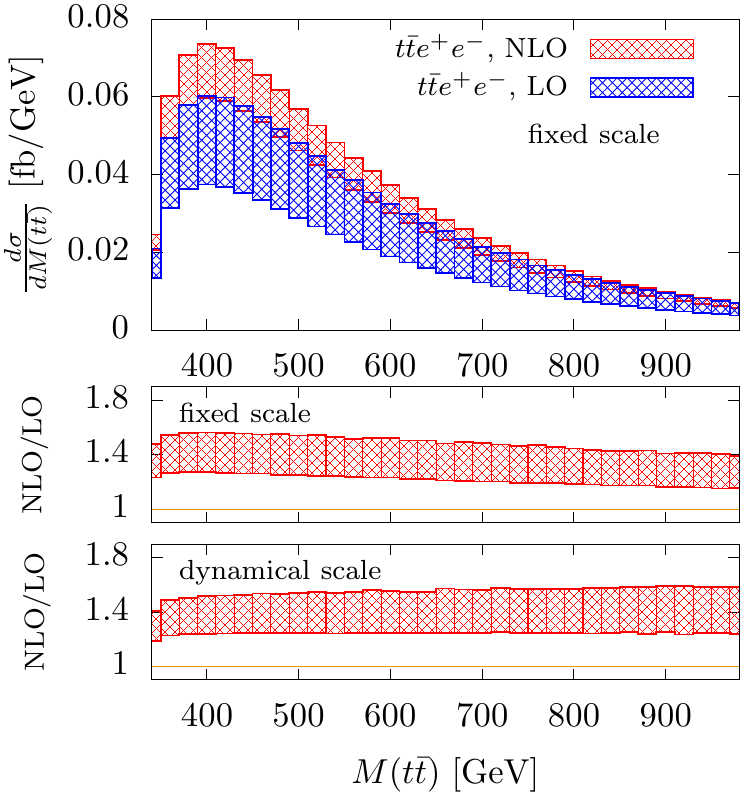}
	\caption{Invariant mass distribution of the $t\bar t$ pair
          in \ttee{} production at LO (blue) and NLO QCD (red)
          accuracy. \label{fig:mtt}}
\end{figure}
%
%%%%%%%%%%
\begin{figure}[htb]
	\centering
	\includegraphics[width=0.48\textwidth]{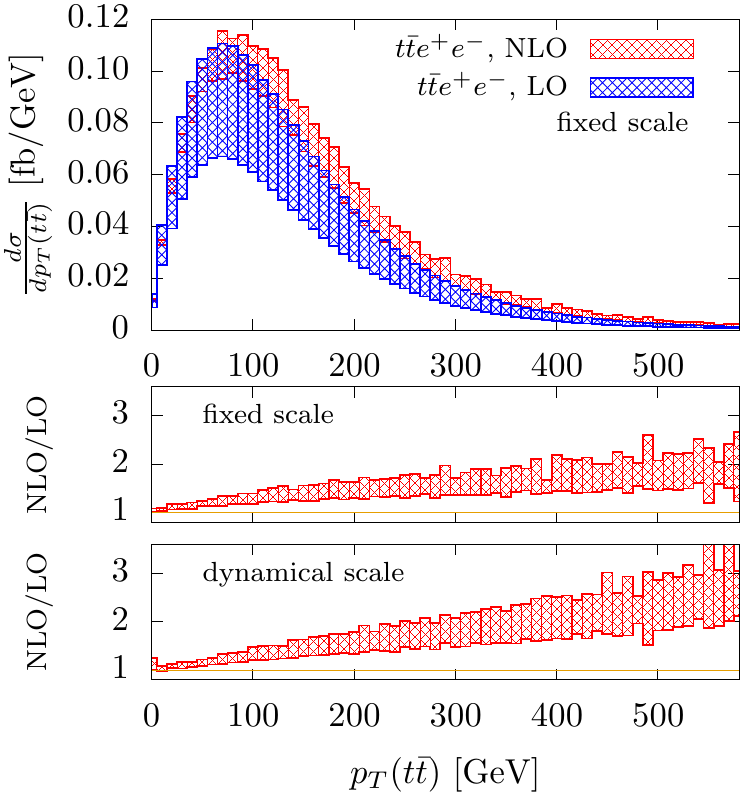}
	\hspace{0.2cm}	
	\includegraphics[width=0.48\textwidth]{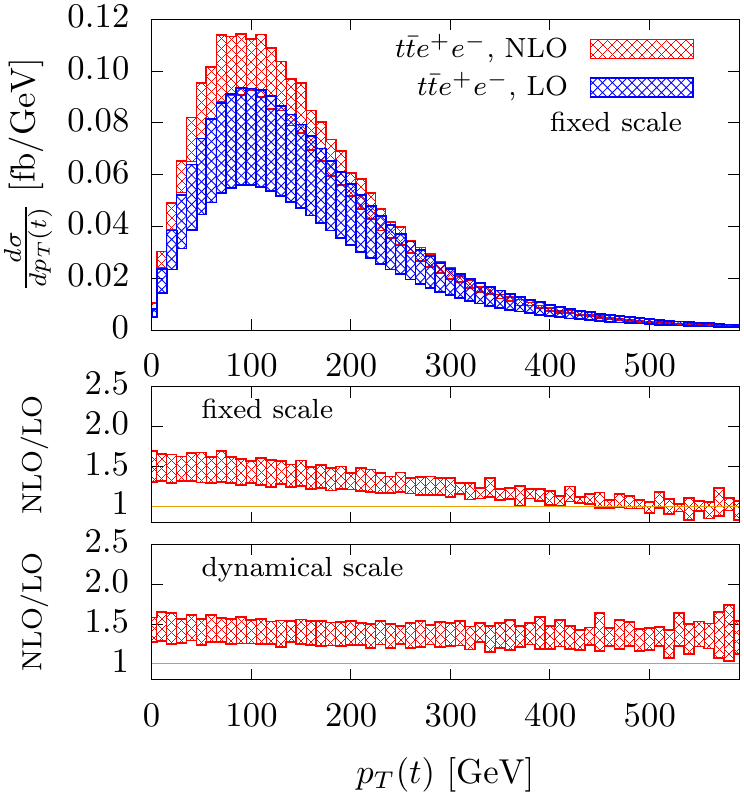}
	\caption{Transverse-momentum distribution of the $t\bar t$ system (left) and 
	of the top quark (right) 
	obtained with our \ttee{} implementations at LO (blue) and NLO
        QCD (red) accuracy. 
	\label{fig:pt-tt}}
\end{figure}
%
%%%%%%%%%%
%
\begin{figure}[htb]
	\centering
	\includegraphics[width=0.48\textwidth]{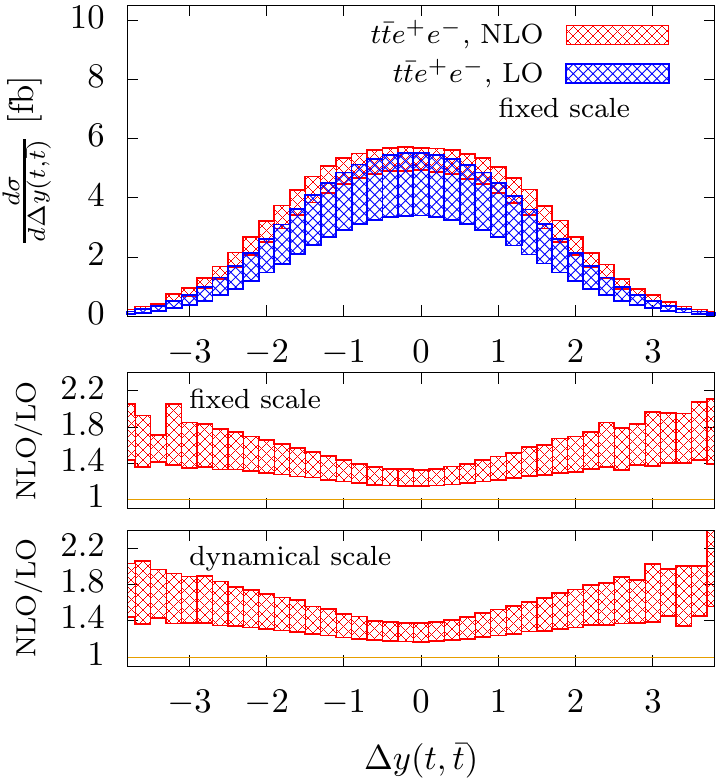}
	\hspace{0.2cm}	
	\includegraphics[width=0.48\textwidth]{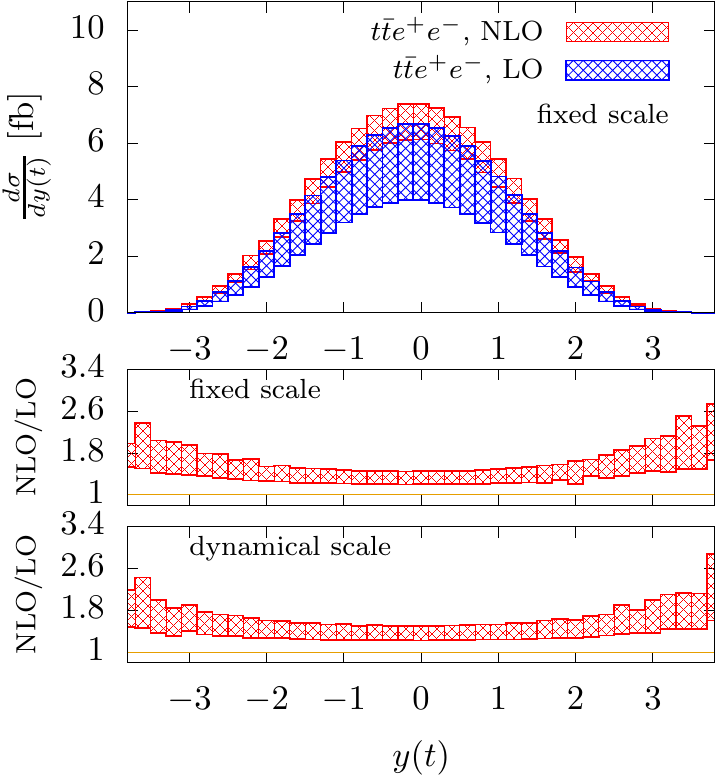}
	\caption{Distribution of the rapidity separation of the top
          and anti-top quark  (left) and of the top quark (right)
          obtained with our \ttee{} implementation at LO and NLO QCD accuracy. 
	\label{fig:eta-tt}}
\end{figure}

\begin{figure}[htb]
	\centering
	\includegraphics[width=0.48\textwidth]{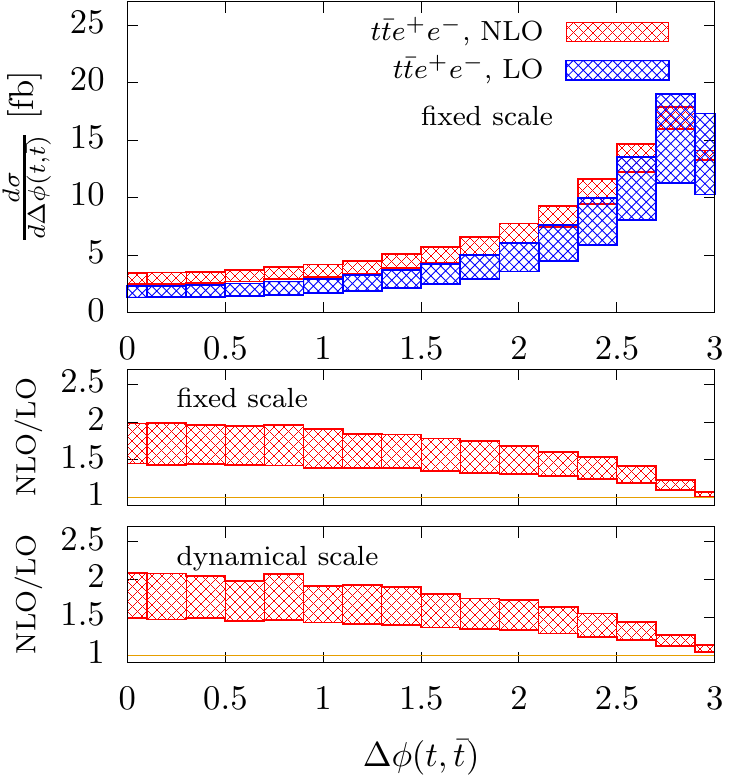}
	\caption{Distribution of the 
	azimuthal angle separation of the top from the anti-top
        obtained with our \ttee{} implementations at LO and NLO QCD accuracy. 
	\label{fig:dphi}}
\end{figure}
%%%%%%%%%%%%%

Figures~\ref{fig:mtt}-\ref{fig:dphi} illustrate the differential
distributions of a variety of observables built from the momenta of
the top quarks at LO and NLO QCD. In each figure the upper panel shows
LO and NLO QCD results for a given distribution using the fixed scale
of Eq.~(\ref{eq:fixed-scale}) and the corresponding uncertainty band
obtained from scale variation, while the middle and lower panels give
the bin-by-bin ratios of NLO QCD versus LO results (i.e. the so called
differential $K$ factor) obtained using the fixed and dynamical scale
choices of Eqs.~(\ref{eq:fixed-scale}) and (\ref{eq:dyn-scale}),
respectively. In general, the middle and lower panels in
Figs.~\ref{fig:mtt}-\ref{fig:dphi} show a very similar behavior
when considering either a fixed or a dynamical scale, even at the
differential level, and well summarize the impact and residual
perturbative uncertainty on the theoretical prediction for \ttll{}
production including NLO QCD corrections.

More specifically, in Fig.~\ref{fig:mtt} we see that the NLO QCD
corrections to the invariant mass distribution of the $t\bar{t}$
system ($M(t\bar{t})$) are quite large and uniformly affect the
corresponding LO results over the extended range considered in
Fig.~\ref{fig:mtt}.  On the other hand, as illustrated by
Fig.~\ref{fig:pt-tt}, the transverse momentum distribution of the
$t\bar{t}$ system ($p_T(t\bar{t})$) receives larger NLO QCD
corrections in the high-$p_T$ regime, where the scale uncertainty is
also larger. A similar behavior has been previously observed in
calculations of $t\bar{t}W^\pm$ production~\cite{Cordero:2021iau}.  In
contrast, the $p_T$ distribution of the top quark ($p_T(t)$) receives
larger NLO QCD corrections in the low $p_T$ region. Both the variation
of the relative NLO QCD correction and of its scale uncertainty
throughout the shown transverse momentum range are less pronounced for
the top quark than for the $t\bar{t}$ system. The case of
$p_T(t\bar{t})$ and $p_T(t)$ distributions is also the only case in
which we notice a mild difference between using a fixed and a
dynamical renormalization and factorization scale. Using a dynamical
scale seems to enhance the effect of NLO QCD corrections in the
high-$p_T(t\bar{t})$ region, while it gives a more uniform $K$ factor in
the case of the $p_T(t)$ distribution.

For the top-quark rapidity ($y(t)$) and the rapidity difference of top
quark and anti-quark ($\Delta y(t,\bar{t})$), shown in
Fig.~\ref{fig:eta-tt}, the relative NLO QCD corrections become larger
for higher values of $y(t)$ and $\Delta y(t,\bar{t})$, and the
associated scale uncertainties also increase with the absolute value
of the rapidity or rapidity difference. Fig.~\ref{fig:dphi} shows the
distribution of the azimuthal-angle difference ($\Delta \Phi(t,\bar{t})$) between the momenta of
the top quark and anti-quark. We observe larger relative NLO QCD
corrections and larger scale uncertainties for both scale choices in
the low $\Delta \phi(t\bar{t})$ region than at high values of
$\Delta \phi(t\bar{t})$.

%%%%%%%%%
\begin{figure}[htb]
	\centering
	\includegraphics[width=0.48\textwidth]{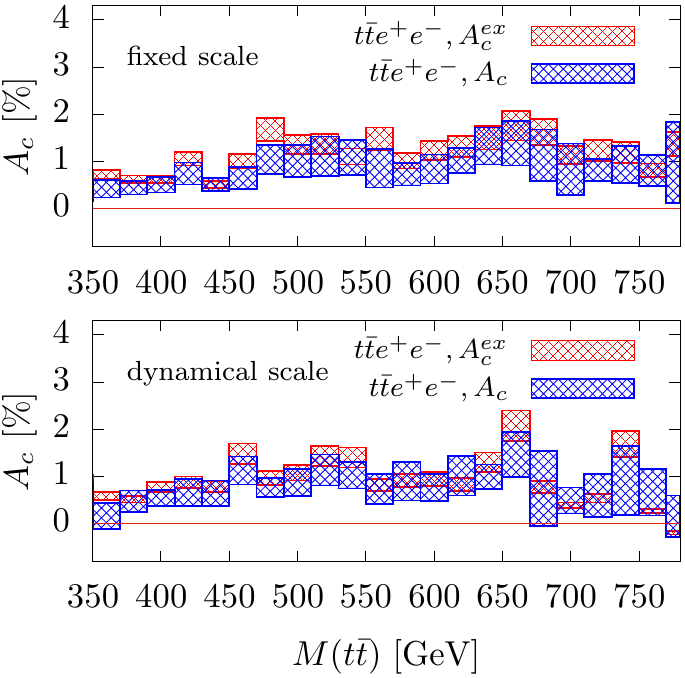}
	\caption{The expanded charge asymmetry $A_c^{ex}(\mtt)$ (red)
          and unexpanded charge asymmetry $A_c(\mtt)$ (blue) for the
          fixed (top) and dynamical scale (bottom) choice as functions
          of the invariant mass of the $t\bar t$ pair in \ttee{}
          production.\label{fig:ac-mtt}}
\end{figure}
Finally, we consider the charge asymmetry of the $t\bar{t}$ system.
In processes involving top-quark pairs, charge asymmetries of the
$\ttbar$ system or its decay products have received considerable
attention (see, e.g., Ref.~\cite{Maltoni:2014zpa} for the case of
\ttw{} production) for constraining effects of physics beyond the SM.
In particular, the charge asymmetry $A_c$ is sensitive to differences
in the absolute values of the rapidity distributions of the top quark
and anti-quark, \beq \dytt = |y_t|-|y_{\bar t}|\,.  \eeq
With $\sigma(\dytt> 0)$ [$\sigma(\dytt< 0)$] denoting the part of the
full cross section within a given set of cuts with positive [negative]
values of $\dytt$, this charge asymmetry is defined as 
\begin{equation}
\label{eq:ac}
A_c = \frac{\sigma(\dytt >0)-\sigma(\dytt < 0)}{\sigma(\dytt> 0)+\sigma(\dytt< 0)}\,,
\end{equation}
where the cross section for each term in the numerator and denominator
is evaluated at NLO QCD accuracy.
While each of the entries in this quantity can be sizable, large
cancellations between cross-section contributions with rapidity
differences of opposite sign result in rather small values of $A_c$,
which makes the numerical calculation of the asymmetry challenging. We
note that a similar trend in the simulation of the charge asymmetry
was reported in the context of $\ttbar$ and $\ttbar$+jet production at
the LHC in Ref.~\cite{Czakon:2017lgo} and Ref.~\cite{Alioli:2011as},
respectively. As discussed in Ref.~\cite{Czakon:2017lgo}, for
instance, the definition of $A_c$ of Eq.~(\ref{eq:ac}) introduces
contributions beyond NLO QCD which are not under control. This can be
avoided by expanding $A_c$ consistently up to ${\cal O}(\alphas)$
which yields 
\begin{equation}
\label{eq:acex}
A_{c}^{ex} = A_c \; \frac{\sigma_\mr{\ttepem}^\mr{NLO}}{\sigma_\mr{\ttepem}^\mr{LO}}\,,
\end{equation}
with the unexpanded charge asymmetry $A_c$ of Eq.~(\ref{eq:ac}) and
the inclusive \ttee{} cross section at NLO QCD ($\sigma_\mr{\ttepem}^\mr{NLO}$) and LO
($\sigma_\mr{\ttepem}^\mr{LO}$) accuracy.  For the unexpanded charge asymmetry of
Eq.~(\ref{eq:ac}) and the expanded charge asymmetry of
Eq.~(\ref{eq:acex}) in \ttee{} production, we find using the fixed
scale of Eq.~(\ref{eq:fixed-scale}):
\begin{eqnarray}
A_c        = 0.84^{+0.28}_{-0.19} \% &\; ,&
A_{c}^{ex} = 1.15^{+0.11}_{-0.18} \% \; , 
\end{eqnarray}
and for the dynamical scale of Eq.~(\ref{eq:dyn-scale}):
\begin{eqnarray}
A_c        = 0.74^{+0.25}_{-0.18} \%  &\; ,&
A_{c}^{ex} = 1.04^{+0.20}_{-0.17} \% \, .
\end{eqnarray}
As illustrated by Fig.~\ref{fig:ac-mtt}, where the charge asymmetry is
shown as a function of the invariant mass of the $\ttbar$~system, the
scale uncertainty is smaller for $A_c^{ex}$ than for $A_c$ for both
the fixed and dynamical scale choices.

\subsection{Assessment of off-shell effects in the $\ell^+\ell^-$~system}
\label{sec:results-off-shell}
After having established the main theoretical uncertainty of NLO QCD
predictions for \ttll{} production, we explore the sensitivity of
\ttll{} results to off-shell and lepton spin-correlation effects
through a comparison of \ttee{} with \ttz{} on-shell production
matched, in the narrow-width approximation, to $Z\rightarrow e^+e^-$
via the \PYTHIA{} decay feature. We consider observables involving the
momenta of the $Z$~boson's decay leptons and compare results obtained
from our \ttee{} and \ttz{} implementations.  In both cases top and
anti-top quark's decays ($t\to \mu^+\nu_\mu b$ and
$\bar t\to \mu^-\bar\nu_{\mu} \bar b$) are treated using the decay
feature of \PYTHIA, since we first aim at isolating effects arising
only from spin-correlations and off-shellness of the leptons not
coming from top-quark decays.

\begin{figure}[t]
	\centering
	\includegraphics[width=0.48\textwidth]{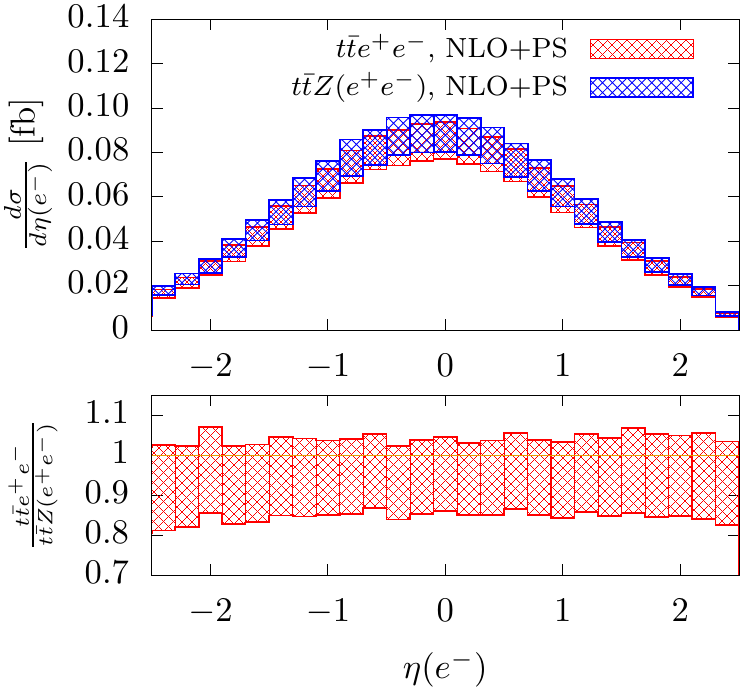}
	\hspace{0.2cm}
	\includegraphics[width=0.458\textwidth]{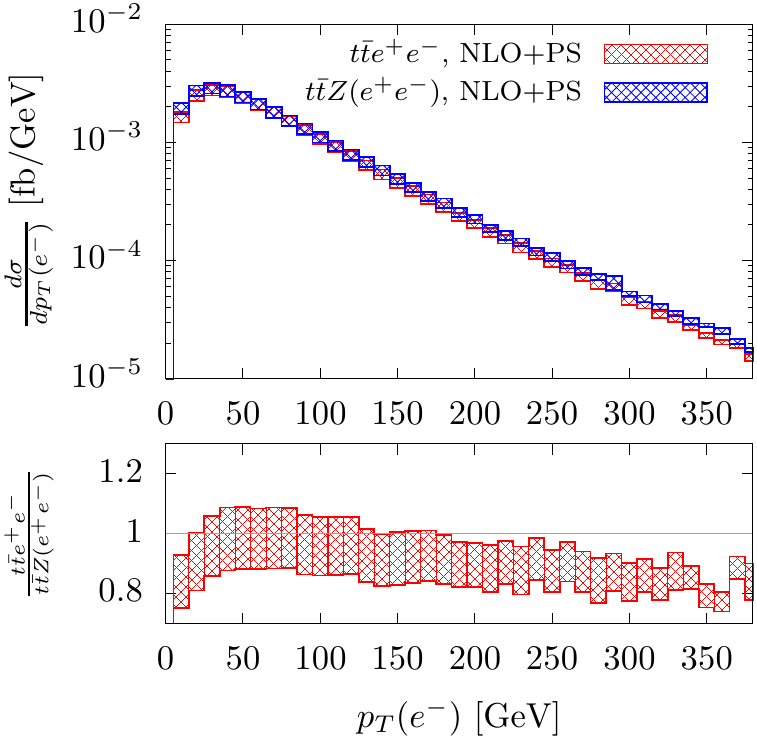}
	\caption{Pseudorapidity (left) and transverse-momentum
          distribution (right) of the electron obtained with our
          \ttz{} (blue) and \ttee{} (red) implementations,
          respectively, at NLO+PS accuracy.\label{fig:pt-eta-e}}
\end{figure}

\begin{figure}[htb]
	\centering
	\includegraphics[width=0.48\textwidth]{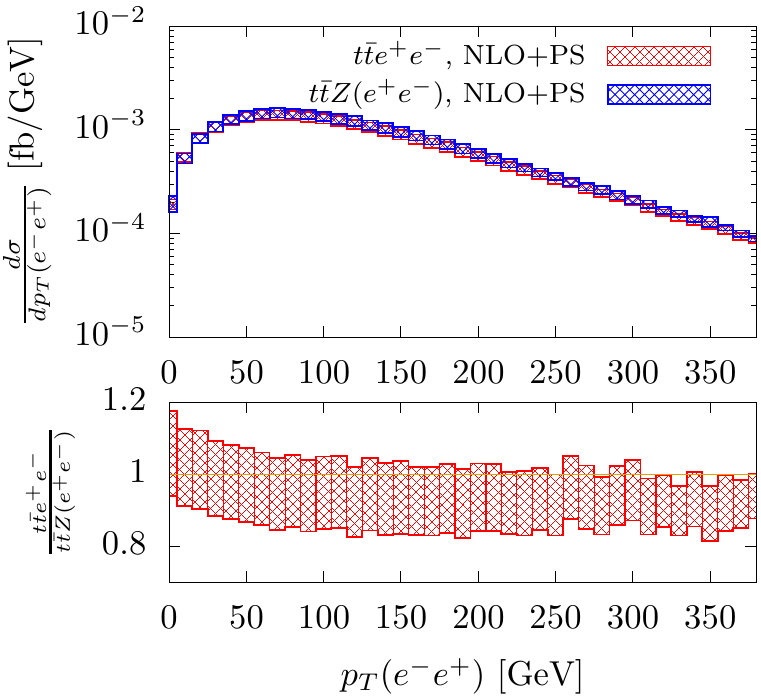}
	\hspace{0.2cm}
	\includegraphics[width=0.48\textwidth]{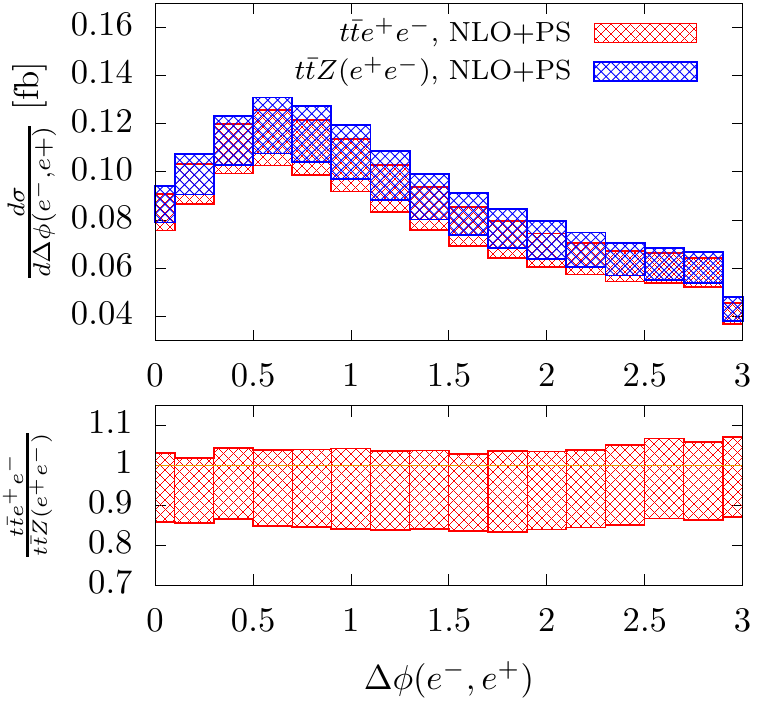}
	\caption{Transverse-momentum distribution of the $\ee$ system
          (left) and azimuthal angle separation of the electron from
          the positron (right) obtained with our \ttz{} (blue) and
          \ttee{} (red) implementations, respectively, at NLO+PS
          accuracy.\label{fig:pt-phi-ee}}
\end{figure}

\begin{figure}[htb]
	\centering
	\includegraphics[width=0.47\textwidth]{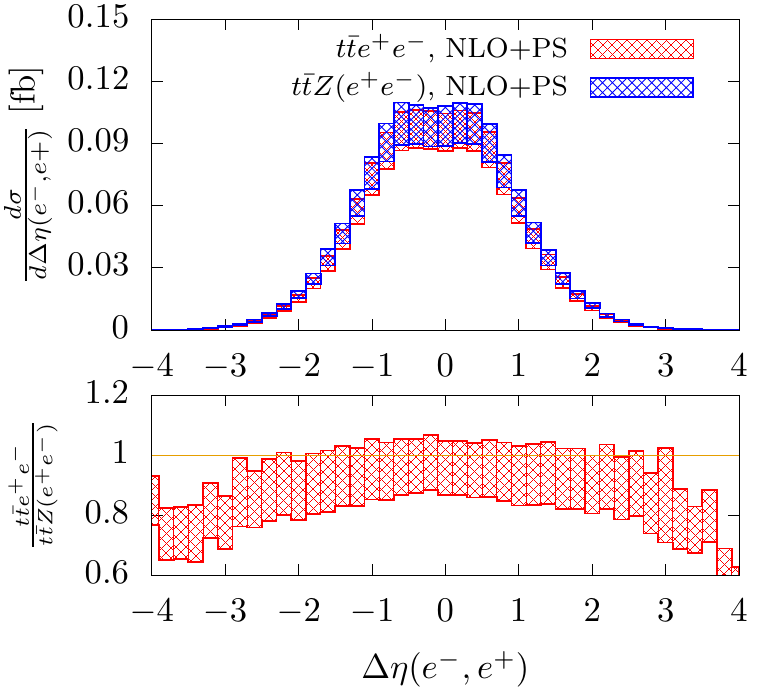}
	\hspace{0.2cm}
	\includegraphics[width=0.47\textwidth]{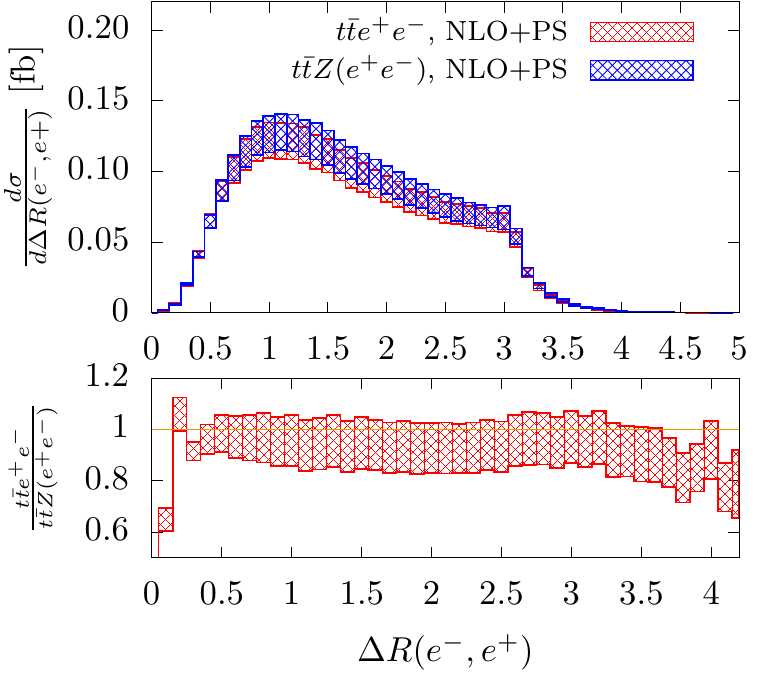}
	\caption{Distribution of the pseudorapidity separation (left)
          and $\Delta R$ separation of the electron from the positron
          (right) obtained with our \ttz{} (blue) and \ttee{} (red)
          implementations, respectively, at NLO+PS
          accuracy.\label{fig:deta-reta-ee}}
\end{figure}

The \ttee{} implementation fully accounts for off-shell contributions
and spin correlations in diagrams where a $Z$~boson decays into an
$\ee$ pair, and also includes diagrams where the lepton pair stems
from a photon rather than a $Z$~boson. While the $\ee$~final-state
system is described with full NLO QCD accuracy in our
\ttee{}~implementation, in the \ttz{}~program decay products of the
$Z$~boson can only be accounted for when event files providing momenta
for the \ttz{}~system are processed with a multi-purpose Monte-Carlo
event generator like \PYTHIA{}, capable to simulate decays of on-shell
particles with no spin-correlation effects at LO accuracy. Such an
approach does not capture off-shell effects or spin correlations in
the decay and, moreover, entirely neglects contributions to the
\ttee{}~final state due to diagrams where a photon is exchanged rather
than a $Z$~boson. A detailed comparison of observables that might be particularly
sensitive to such effects reveals that the two implementations differ
in specific regions of phase space, as illustrated in
Figs.~\ref{fig:pt-eta-e}-\ref{fig:deta-reta-ee}. It is interesting to
notice that such effects are visible even if we restrict the fiducial
volume to a window region around the mass of the $Z$ boson by using
the cut of Eq.~(\ref{eq:mee-cut}).

The following discussion is based on NLO+PS results obtained with
\PYTHIA. The electron cuts of Eqs.~(\ref{eq:ptl-cut}) and
(\ref{eq:mee-cut}) are applied.  
The corresponding total cross sections
including scale uncertainties for the fixed scale choice of
Eq.~(\ref{eq:fixed-scale}) are
\begin{eqnarray}
\sigma_\mr{t\bar tZ(\ee)}^\mr{NLO+PS} & = & 0.274^{+0.025}_{-0.030}  \, \mr{fb}\;, \\ \nonumber
\sigma_\mr{\ttepem}^\mr{NLO+PS}          & = & 0.261^{+0.024}_{-0.028}  \, \mr{fb} \; ,
\end{eqnarray}
and for the dynamical scale of Eq.~(\ref{eq:dyn-scale}) we find 
\begin{eqnarray}
\sigma_\mr{t\bar tZ(\ee)}^\mr{NLO+PS} & = & 0.275^{+0.028}_{-0.031}  \, \mr{fb}\;, \\ \nonumber
\sigma_\mr{\ttepem}^\mr{NLO+PS}          & = & 0.264^{+0.027}_{-0.030}  \, \mr{fb} \; .
\end{eqnarray}
We present results for the
pseudorapidity ($\eta(e^-)$) and transverse-momentum ($p_T(e^-)$)
distributions of the electron in Fig.~\ref{fig:pt-eta-e}, the
transverse-momentum ($p_T(e^-e^+)$) and azimuthal-angle separation
($\Delta\phi(e^-,e^+)$) of the $\ee$ system in
Fig.~\ref{fig:pt-phi-ee}, as well as its pseudorapidity separation
($\Delta\eta(e^-,e^+)$) and $\Delta R$ separation ($\Delta R(e^-,e^+)$)
in Fig.~\ref{fig:deta-reta-ee}, where $\Delta R(\ell^-,\ell^+)$ is defined in terms
of $\Delta \phi$ and $\Delta\eta$ of a generic $\ell^-\ell^+$ pair as:
\begin{equation}
\label{eq:deltaR-ll}
  \Delta R(\ell^-,\ell^+) = \sqrt{\Delta \phi^2(\ell^-,\ell^+)+\Delta \eta^2(\ell^-,\ell^+)} \;.
\end{equation}
The differential distributions obtained with our \ttee{} and \ttz{}
implementations and shown in
Figs.~\ref{fig:pt-eta-e}-\ref{fig:deta-reta-ee} agree within the scale
uncertainty band in most regions of phase space. However, there are
substantial effects of the order of 10-20\%, visible outside the scale
uncertainty band, most notably in the high-$p_T$ region of the
electron/positron transverse-momentum distribution as illustrated in
Fig.~\ref{fig:pt-eta-e} for the case of $p_T(e^-)$, in the large
absolute-value region of the pseudorapidity difference
$\Delta\eta(e^-,e^+)$
%as shown in Fig.~\ref{fig:pt-phi-ee}, and in
%corresponding regions of the $\Delta R (e^-,e^+)$ distance as one can
%see in Fig.~\ref{fig:deta-reta-ee}.
and in corresponding regions of the $\Delta R (e^-,e^+)$ distance as one can see in Fig.~\ref{fig:deta-reta-ee}.
%%%%%%%%%%
%\begin{figure}[htb]
%%%%%%%%%%
%%%%%%%%%%
%%%%%%%%%%
%%%%%%%%%%
%%%%%%%%%%

\subsection{Assessment of spin correlations in top decays}
\label{sec:results-top-decays}
Next, we would like to assess the effect of modeling on-shell
top-quark decays, in the narrow-width approximation, including
tree-level spin correlations as obtained in our \ttee{} implementation
following the method of Ref.~\cite{Frixione:2007zp} versus a more
approximate treatment where decays of the top quarks are simulated by
the decay feature of \PYTHIA{} and no spin correlations are retained.
For simplicity, in the following these two simulations are referred to
as {\em with spin correlations} and {\em without spin correlations},
although in both cases spin correlations in the $\ee$ system are fully
taken into account. We remind the reader that, as explained in
Sec.~\ref{sec:implementation}, both options are available in our
\POWHEGBOX{} implementation. In the first case, at the Les-Houches
event (LHE) level before parton shower, the program generates fully
decayed events, i.e.
$(t\to \mu^+\nu_\mu b)(\bar t\to \mu^-\bar\nu_{\mu} \bar b)\, e^+e^-$,
while in the second case only \ttee{} events are generated.
%%%%%%%%%%
\begin{figure}[htb]
	\centering
		\includegraphics[width=0.48\textwidth]{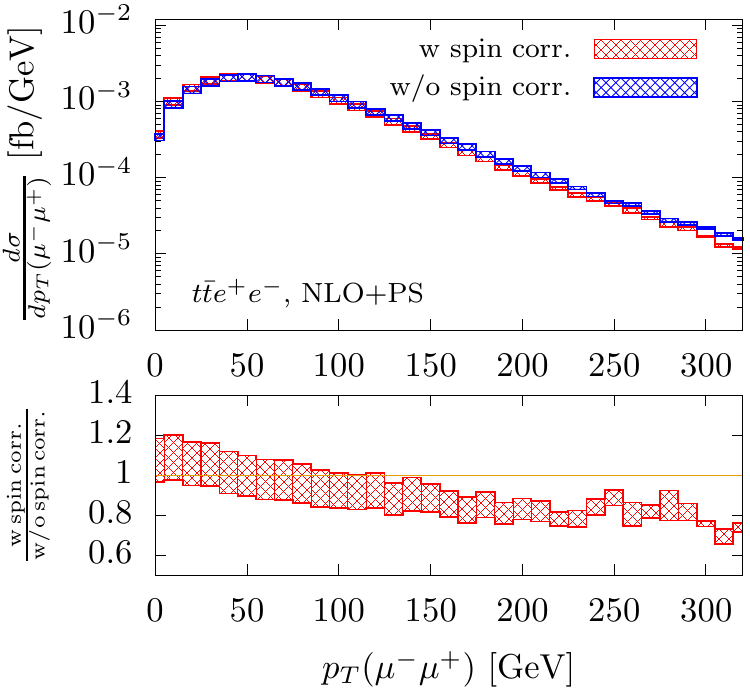}
		\hspace{0.2cm}
		\includegraphics[width=0.48\textwidth]{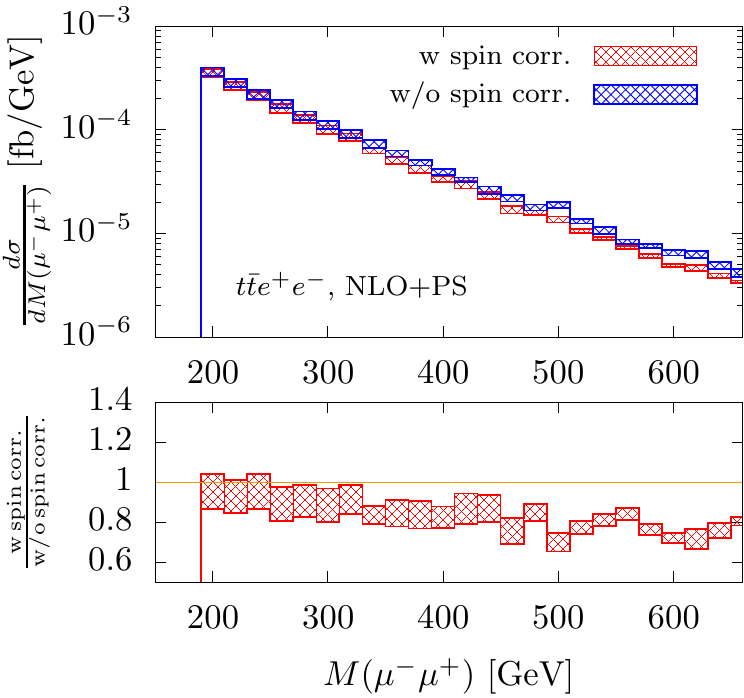}
	\caption{
	\label{fig:pt-m-mumu}
        Transverse-momentum (left) and invariant-mass distribution of
        the $\mumu$ system (right) in the $pp\to \ttzdec$ process at
        NLO+PS accuracy. Correlations in the top-quark decays are
        either taken into account at the level of event generation
        (red) or provided by \PYTHIA~(blue).}
\end{figure}
%%%%%%%%%%
%%%%%%%%%%
\begin{figure}[htb]
	\centering
	\includegraphics[width=0.47\textwidth]{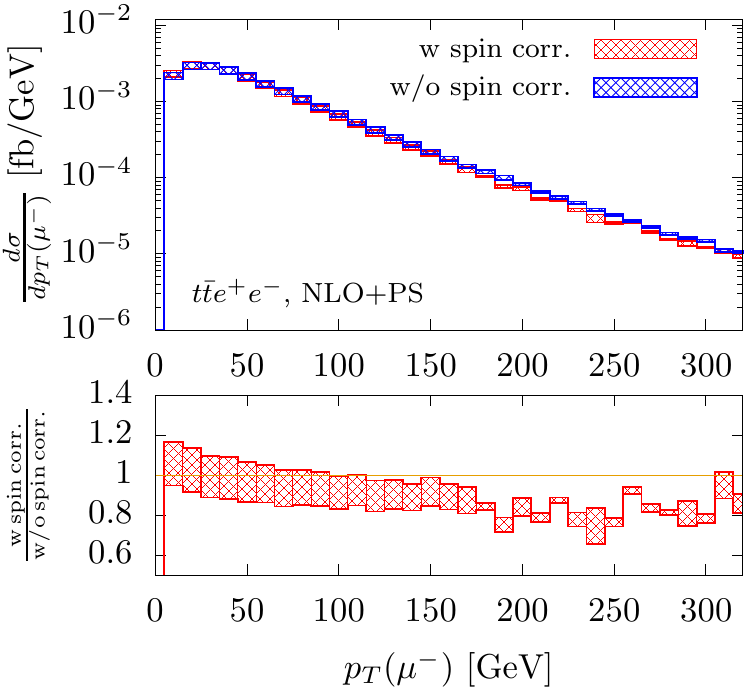}
	\hspace{0.2cm}
	\includegraphics[width=0.47\textwidth]{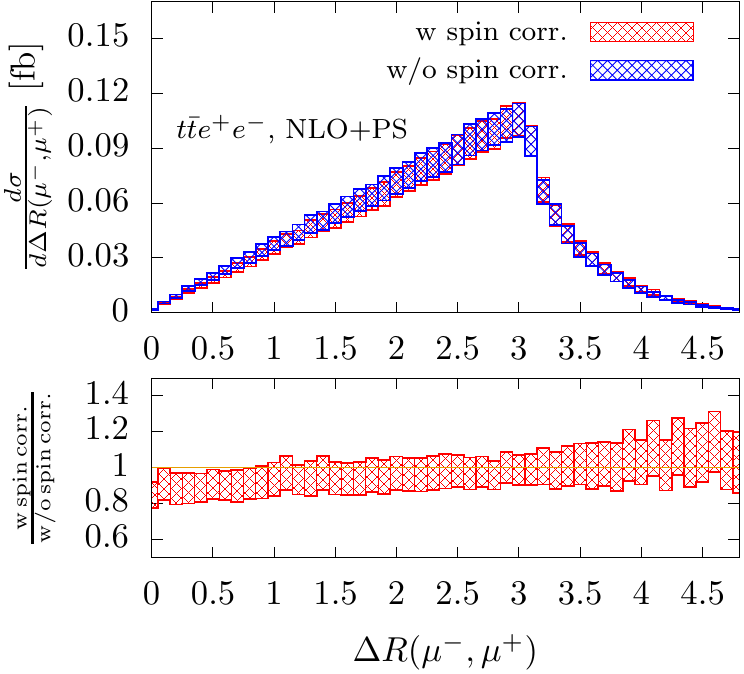}
	\caption{
	\label{fig:pt-r-mu}
        Transverse-momentum distribution of the muon (left) and
        $\Delta R$~separation of the $\mu^-$ from the $\mu^+$ (right)
        in the $pp\to \ttzdec$ process at NLO+PS
        accuracy. Correlations in the top-quark decays are either
        taken into account at the level of event generation (red) or
        provided by \PYTHIA~(blue).}
\end{figure}
%%%%%%%%%%

For the results shown in this section, the muon cuts of
Eq.~(\ref{eq:muon-cuts}) are applied in addition to the electron cuts
of Eqs.~(\ref{eq:ptl-cut}) and (\ref{eq:mee-cut}).  The corresponding
total cross sections including scale uncertainties for the fixed scale
choice of Eq.~(\ref{eq:fixed-scale}) are
\begin{eqnarray}
\mr{with\; spin \; corr.} & : & \sigma_\mr{\ttepem}^\mr{NLO+PS} =
                                0.199^{+0.030}_{-0.011}  \, \mr{fb} \;,\\ \nonumber
\mr{w/o \; spin \; corr.} & : & \sigma_\mr{\ttepem}^\mr{NLO+PS} = 0.214^{+0.019}_{-0.023}  \, \mr{fb} \; ,
\end{eqnarray}
while for the dynamical scale of Eq.~(\ref{eq:dyn-scale}) we find 
\begin{eqnarray}
\mr{with\; spin \; corr.} & : & \sigma_\mr{\ttepem}^\mr{NLO+PS} = 0.216^{+0.021}_{-0.024}  \, \mr{fb} \,,\\ \nonumber
\mr{w/o \; spin \; corr.} & : & \sigma_\mr{\ttepem}^\mr{NLO+PS} = 0.213^{+0.020}_{-0.023}  \, \mr{fb} \; .
\end{eqnarray}
As expected, effects on the total cross section are small while
they are more relevant in specific regions of kinematic
distributions. Indeed, a more detailed comparison of the two approaches
in terms of distributions of kinematic observables built from the
momenta of the top-quark and $Z$-boson decay products reveals the
presence of non negligible effects, in particular for observables of
the ($\mu^+\mu^-$) system.

A clear example is illustrated in Fig.~\ref{fig:pt-m-mumu}, where we
show the transverse momentum ($p_T(\mu^-\mu^+)$) and the invariant
mass ($M(\mu^-\mu^+)$) distributions of the $\mumu$ system. The tails
of both of these distributions are considerably lower, by roughly
10-20\%, for the predictions that take into account spin correlations
in the top-quark decays and the effect is clearly visible on top of
the renormalization and factorization scale uncertainty considered in
our study. A similar effect can be observed in the transverse-momentum
distribution of the muons ($p_T(\mu^\pm)$), as depicted in
Fig.~\ref{fig:pt-r-mu} for the case of $p_T(\mu^-)$. Milder yet visible effects are
observed in angular distributions such as the $\Delta R$~separation of
$\mu^-$ from $\mu^+$  defined
%in terms of their azimuthal angle and pseudorapidity separations
as in Eq.~(\ref{eq:deltaR-ll}), 
%\begin{equation}
%\Delta R(\mu^+,\mu^-) = \sqrt{\Delta \phi^2(\mu^+,\mu^-)+\Delta \eta^2(\mu^+,\mu^-)} \; , 
%\end{equation}
which is dominated by contributions of low or moderate transverse momentum.

\begin{figure}[htb]
	\centering
		\includegraphics[width=0.48\textwidth]{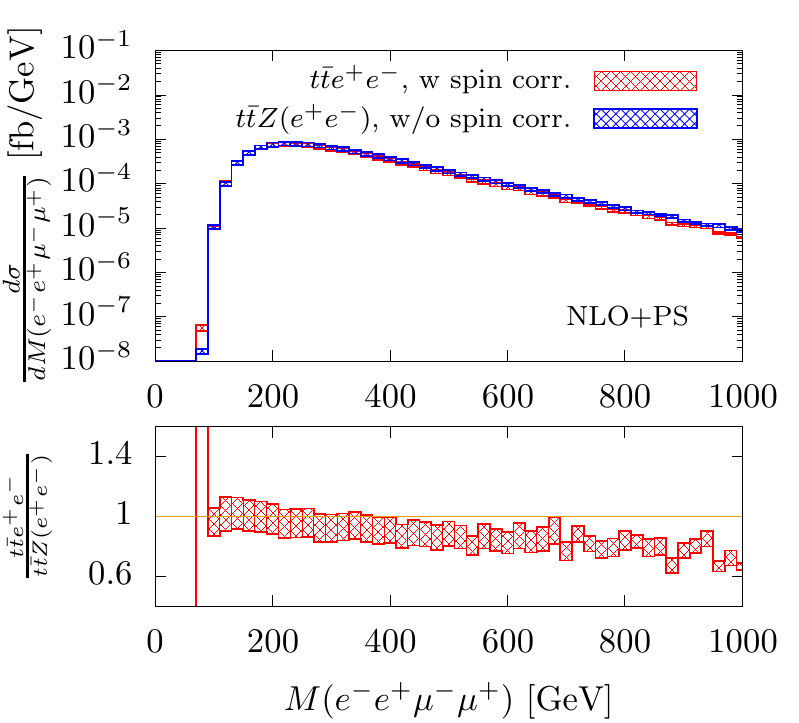}
		\hspace{0.2cm}
		\includegraphics[width=0.48\textwidth]{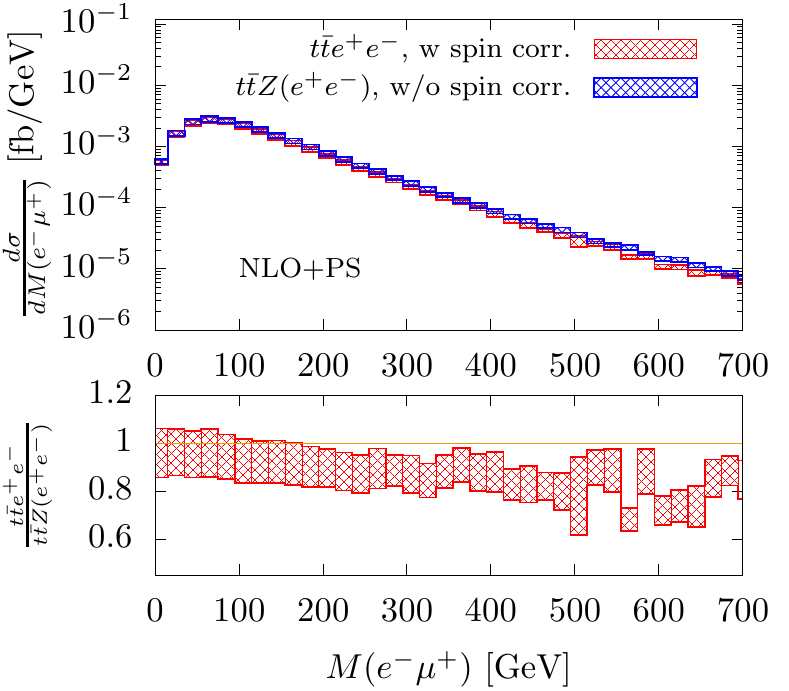}
	\caption{
	\label{fig:mass-me}
        Invariant-mass distribution of the $\ee\mumu$ system (left)
        and of the $e^- \mu^+$ system (right) in the $pp\to \ttzdec$
        and $pp\to \ttzdecz$ processes at NLO+PS
        accuracy. Correlations in the top-quark and $Z$-boson decays are either
        taken into account at the level of event generation (red) or
        provided by \PYTHIA~(blue).}
\end{figure}

\begin{figure}[htb]
	\centering
		\includegraphics[width=0.48\textwidth]{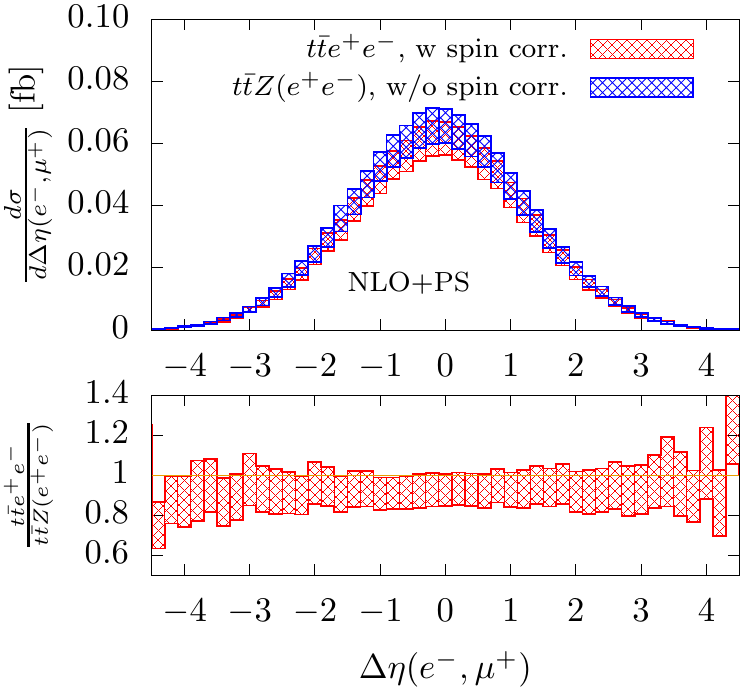}
		\hspace{0.2cm}
		\includegraphics[width=0.48\textwidth]{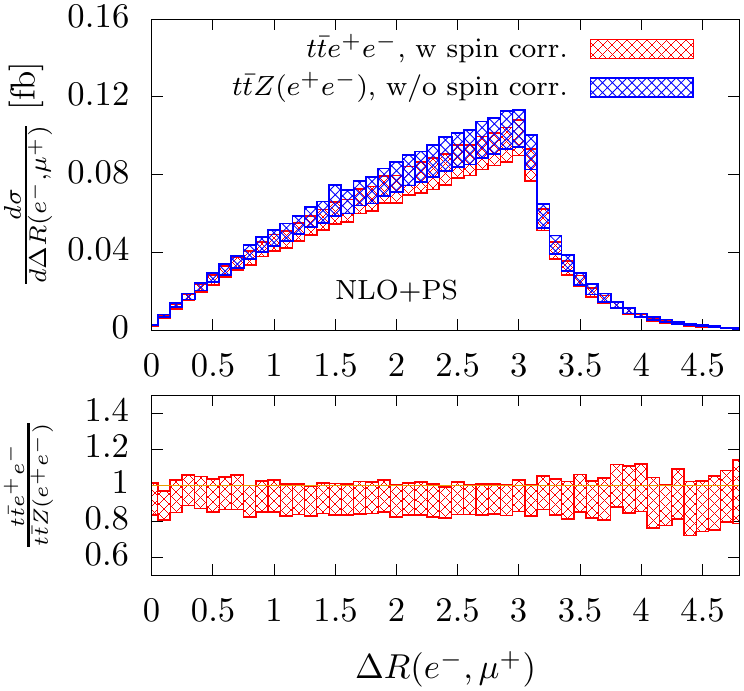}
                \caption{ Distribution of the pseudorapidity
                  separation (left) and $\Delta R$ separation of the
                  electron from the $\mu^+$ (right) in the
                  $pp\to \ttzdec$ and $pp\to \ttzdecz$ processes at
                  NLO+PS accuracy. Correlations in the top-quark and $Z$-boson
                  decays are either taken into account at the level of
                  event generation (red) or provided by
                  \PYTHIA~(blue).
\label{fig:deta-reta-em-FS}}
\end{figure}

Having access to the momenta of the muons stemming from the top-quark
decays as well as to the $\ee$ system, we can also explore
correlations between the various final-state leptons in the
$pp\to \ttzdec$ process. As an example, we present in
Fig.~\ref{fig:mass-me} the invariant mass distributions of the
$\ee\mumu$ and the $e^-\mu^+$ systems, respectively, and we compare
results obtained using the \ttee{} implementation with spin-correlated top decays versus those
obtained from the corresponding implementation of \ttz{} with on-shell
$Z\rightarrow e^+e^-$ decays via \PYTHIA{} default routines, i.e. without spin correlations.  Both
distributions show 10-20\% off-shell and spin-correlation effects for high invariant
masses, while we do not find appreciable effects in the case of
angular distributions of the $e^\pm\mu^\mp$ systems, such as
pseudorapidity differences and $\Delta R$ separations, as illustrated
in Fig.~\ref{fig:deta-reta-em-FS} for the $e^-\mu^+$ case.
%%%%%%%%%%
%%%%%%%%%%
%%%%%%%%%%
%%%%%%%%%%

%%%%%%%%%%%%%%%%%%%%%%%%%%%%%%%%%%%%%%%%%%%
\section{Summary and conclusions}\label{sec:conclusion}
%%%%%%%%%%%%%%%%%%%%%%%%%%%%%%%%%%%%%%%%%%%
%
In this article we present results for the production of \ttll{} at
the LHC with $\sqrt{s}=13$~TeV including NLO QCD corrections matched
to parton showers via the \POWHEG{} method. The option for simulating
decays of the top quarks in a narrow-width approximation including
tree-level spin correlations is provided.

We explicitly study the impact of off-shell effects by comparing
results for \ttll{} production versus results obtained from \ttz{}
on-shell production matched to the $Z\rightarrow \ell^+\ell^-$ decay
in a narrow-width approximation. For illustration purposes we specify
our discussion to the \ttee{} case. We find general compatibility
between the two implementations, but we also notice sizable effects at
the 10-20\% level, clearly visible in the tails of the leptons'
transverse-momentum distributions, as well as in the transverse
momentum and pseudorapity distributions of the $\ell^+\ell^-$ system.

Furthermore, we investigare the effect of considering approximate
spin-correlations among the leptons from top quark and antiquark
decays, as well as among all leptons in the final-state signatures
(from both $Z$-boson and top/antitop decays). We illustrate our
results for the specific case of \ttee{} with $t\rightarrow
\mu^+\nu_\mu b$ and
$\bar{t}\rightarrow\mu^-\bar{\nu}_\mu\bar{b}$. Although small at the
level of total cross sections, these effects can reach 10-20\% in
tails of distributions for transverse momenta and invariant masses of
the $Z$-boson and the top-quark and top-antiquark decay products.

Our implementations of \ttll{} and \ttz{} in \POWHEGBOX{} allow to
study any other \ttll{} signature, provided due care is taken to
adjust the selection of the desired final state signature. Hence, our
study and the tools developed in its context represent substantial
progress towards a full description of production and decay of a
\ttll{} system at hadron colliders, including NLO QCD corrections
matched with parton shower, and will allow a more adequate assessment
of the theoretical uncertainty stemming from the modeling of the kind
of complex final states that are considered in LHC measurements of SM
properties and searches of new physics.  In particular they will allow
a more accurate exploration of the properties of the top quark and its
interactions with EW gauge bosons and the Higgs boson, where the
\ttz{} signatures play a major role either as signal or background,
respectively. With this respect, the kind of effects described in this
paper can be of great relevance since new physics effects tend to
appear in high-mass and high-momentum tails of distributions that will
become more and more statistically significant with the upcoming
high-luminosity runs of the LHC.

Both the \ttll{} and \ttz{} implementations that we developed for this
study will be made available from the website of the \POWHEGBOX{}
project,
\href{http://powhegbox.mib.infn.it/}{http://powhegbox.mib.infn.it/}.

%%%%%%%%%%%%%%%%%%%%%%%%%%%%%%%%%%%%%%%%%%%
\section*{Acknowledgements}
%%%%%%%%%%%%%%%%%%%%%%%%%%%%%%%%%%%%%%%%%%%
The authors would like to thank M. Kraus and C. Oleari for valuable
discussions and help with several aspects of the \POWHEGBOX{}
implementation. M.~G., B.~J., and S.~L. would like to thank
J.~Scheller and C.~Borschensky for helpful conversations.
L.R. would like to thank D. Figueroa and S. Quackenbush for their support in optimizing the use of NLOX for this project.

The work of L.~R. is supported in part by the U.S. Department of
Energy under grant DE-SC0010102. The work of D.~W. is supported in
part by the U.S. National Science Foundation under award
no.~PHY-1719690 and no.~PHY-2014021. Part of this work was performed
on the high-performance computing resource bwForCluster NEMO with
support by the state of Baden-W\"urttemberg through bwHPC and the
German Research Foundation (DFG) through grant no INST 39/963-1 FUGG.

\clearpage
%\bibliographystyle{JHEP}
%\bibliography{ttz}

\providecommand{\href}[2]{#2}\begingroup\raggedright\endgroup

\end{document}